# Learning more from the Lorentz transformations

## Stefan Popescu[1] and Bernhard Rothenstein[2]

1) Siemens AG, Erlangen, Germany, stefan.popescu@siemens.com
2) Politehnica University of Timisoara, Physics Department, Timisoara, Romania, brothenstein@gmail.com

***Abstract***. *Admitting the validity of Lorentz transformations for the space as time coordinates of the same event we derive their differential form in order to underline the correct prerequisites for the application of time and length contraction or dilation effects. Furthermore we quantify the simultaneity error occurring in the relativity theory. Having done this, we analyse the root cause of these effects and identify it with a finite phase velocity associated with the moving frame. We define this phase velocity by analogy to the de Broglie wave associated with a moving particle. Based on this construct we demonstrate that the phase of the de Broglie waves further extended for stationary particles is a relativistic invariant being the same for all corresponding observers. Also the phase of the electromagnetic waves transporting energy at light speed is a relativistic invariant. Therefore the universe and its matter / energy may be seen as a superposition of waves propagating such that their phase is the same for all corresponding observers. The wave phase may replace the time and space as invariant and universal references.*

## 1. Introduction

The Lorenz transformations – LT are very useful tools to solve a lot of problems associated with the special behavior of space-time coordinates of observers moving at high-speed as predicted by Einstein's special relativity theory. However, many authors classify them as not transparent from a physical point of view and avoid using them directly. Some others prefer to derive these equations starting with alternative theoretical constructs or particular assumptions that occur as dogmatic as well for the beginner. For those familiar with the special relativity theory it is clear that all equations and laws in this field are connected with each other by a complex network of inference paths, such that one can be derived under the assumption of others being valid. Starting with the Einstein's famous one[1] a lot of complementary derivations for the Lorenz transformations are known[2] and therefore we conclude that they provide a solid basis for further investigations. Therefore in this paper we start by admitting that the LT are well-funded and we investigate their particular implications that would support a better understanding and easier acceptance of relativistic effects.

## 2. The Lorenz transformations

These transformations relates the space and time coordinates of the same single event $E(x,t)$ / $E'(x',t')$ as seen by two corresponding observers $O$ and $O'$ located just where the event takes place. The two observers are stationary each in other inertial reference frame – $O$ in frame K and respectively $O'$ in a frame K' that moves relative to frame K. These relations are:

$$\text{a) } t' = \gamma \cdot \left( t - \frac{V}{c^2} x \right) \qquad \text{b) } x' = \gamma \cdot (x - Vt) \qquad (1)$$

with $\gamma = \dfrac{1}{\sqrt{1-\beta^2}}$ being the Lorentz factor and $\beta = \dfrac{V}{c}$ the relative frame speed.

Expressed for the time intervals and distances separating two events $E_1(x_1,t_1)$ / $E_1'(x_1',t_1')$ and $E_2(x_2,t_2)$ / $E_2'(x_2',t_2')$ these relations become:

$$\text{a) } \Delta t' = \gamma \left( \Delta t - \frac{V}{c^2} \cdot \Delta x \right) \qquad \text{b) } \Delta x' = \gamma (\Delta x - V\Delta t) \qquad (2)$$

with:

$$\begin{array}{ll} \Delta x = x_2 - x_1 & \Delta x' = x_2' - x_1' \\ \Delta t = t_2 - t_1 & \Delta t' = t_2' - t_1' \end{array} \qquad (3)$$

We obtain the differential form of the Lorentz equations (2) by applying (1) separately for each event, subtracting the resulting equations and using the notations (3) to get the final form of (2).

The inverse Lorenz transformation for the space and time coordinates of the single event $E(x,t)$ / $E'(x',t')$ results as the simple algebraic solution of (1):

$$\text{a) } t = \gamma \left( t' + \frac{V}{c^2} \cdot x' \right) \qquad \text{b) } x = \gamma (x' + Vt') \qquad (4)$$

Expressed differentially for the time intervals and distances separating two events these relations become:

$$\text{a) } \Delta t = \gamma \left( \Delta t' + \frac{V}{c^2} \cdot \Delta x' \right) \qquad \text{b) } \Delta x = \gamma (\Delta x' + V\Delta t') \qquad (5)$$

### 3. The time dilation effect

$$\Delta t = \frac{\Delta t'}{\sqrt{1-\beta^2}} \qquad (6)$$

May be obtained as a particular case of (5a) with $\Delta x' = 0$ thus applicable for two events that occurs at same position in frame K' but at different time instants. Therefore this effect involves a single observer $O'(x' = x_1' = x_2')$ stationary in frame K' and two synchronised observers $O_1(x_1)$ and $O_2(x_2)$ stationary in frame K. One classical example is the moving light clock below.

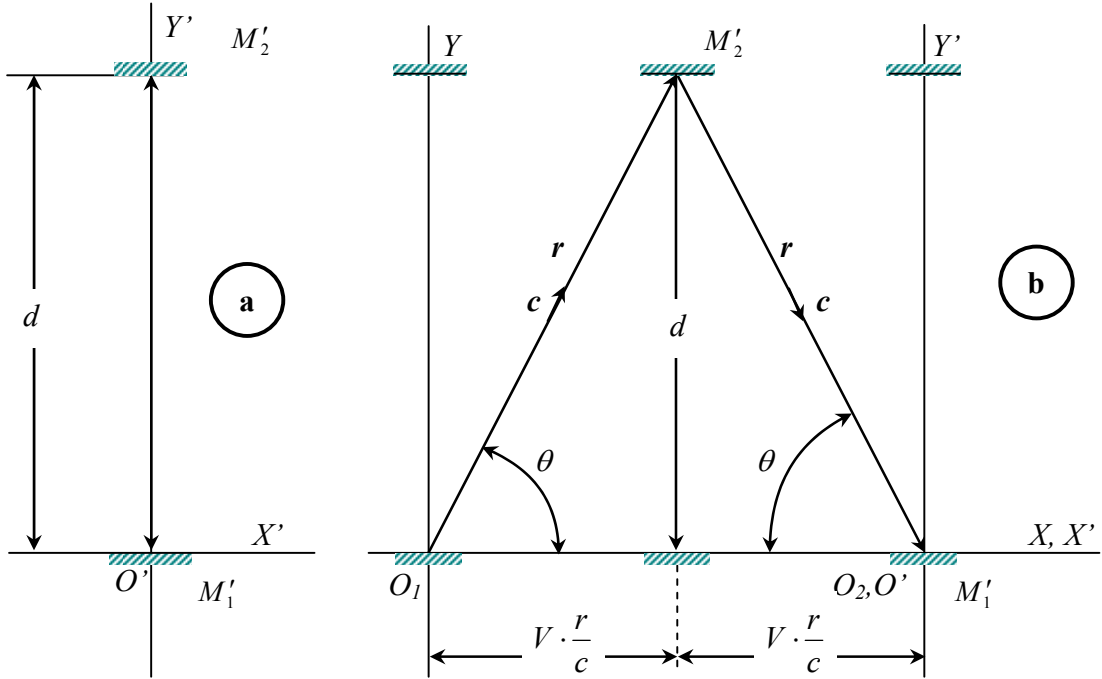

**Figure 1**. *The moving light clock illustrating the time dilation effect*

Figure 1 above depicts a particular example for the correct application of the time dilation effect as derived from the Lorenz transformations. A light clock in frame K' consists of two parallel mirrors and a light signal that bounces cyclically between the two mirrors. As explained above we identify the correct prerequisites for using the time dilation law:

- A single observer $O'$ stationary in frame K' – that measures the period of these oscillations as $T' = \dfrac{2d}{c}$ (figure 1a).

- Two synchronised observers $O_1$ and $O_2$ organised as a team in frame K – that measure the same period above as $T = \dfrac{2r}{c}$.

By applying the Pythagoras's theorem to figure 1b we obtain $d = r\sqrt{1 - \dfrac{V^2}{c^2}}$ and therefore $T = \dfrac{T'}{\sqrt{1 - \dfrac{V^2}{c^2}}}$ i.e. the time dilation law. Under these circumstances we obtain an additional result - from (5a) with $\Delta x' = 0$ we have $\Delta x = \dfrac{V \Delta t'}{\sqrt{1 - \beta^2}} = V \Delta t$ and using (6) we get $V = \dfrac{\Delta x}{\Delta t}$ which is the obvious relation used by observers in K to calculate the speed of the observer commoving with the frame K'.

We stress again that the important prerequisites for the correct application of this effect can be simple obtained from the differential Lorenz equations (2) and (5). For example with $\Delta x = 0$ in (2a) we get the complementary relation:
$$\Delta t = \Delta t' \sqrt{1-\beta^2} \qquad (7)$$
which is actually a time contraction effect when seen from frame K. It involves two events occurring at different time instants but at the same position in frame K as seen by a single observer stationary in frame K and two observers stationary in frame K'. Many authors prefer to use the concept of proper time for the time interval measured by the single observer in order to differentiate it from the time interval measured by two synchronised observers working as a team. In this perception the proper time interval is always shorter than any other.

### 4. The length contraction effect
$$\Delta x = \Delta x' \sqrt{1-\beta^2} \qquad (8)$$

In many text books this effect is presented as a contraction of moving objects when perceived by observers from the stationary frame. Being unsatisfied with such explanation we turn to our differential Lorentz equations and we try to learn more from them. We may obtain (8) as a particular case of (2b) with $\Delta t = 0$ and $\Delta x \neq 0$ thus applicable for two simultaneously but distant events in frame K. The prerequisites for the correct application of this law involves two distant and synchronised observers $O_1(x_1)$ and $O_2(x_2)$ stationary in frame K and two synchronised observers $O'_1(x'_1)$ and $O'_2(x'_2)$ stationary in frame K'. Reformatting (8) in the equivalent form $\Delta x' = \Delta x / \sqrt{1-\beta^2}$ we could draw the straight conclusion: the distance between two events that are simultaneous in the stationary frame occurs longer for the moving observers. Therefore we deal actually with a length dilation effect. However, the two simultaneous events in the stationary frame don't occur to be also simultaneous for the moving observers. From (2a) with $\Delta t = 0$ and using (8) we obtain that:
$$\Delta t' = \gamma \left( -\frac{V}{c^2} \cdot \Delta x \right) = -\frac{V}{c^2} \Delta x' \qquad (9)$$
Therefore the deviation from simultaneity depends on the relative frame speed $V$ and the apparent distance between the events.

Now suppose that two events occurs simultaneously for two distant observers stationary in the moving frame K' ($\Delta t' = 0$ and $\Delta x' \neq 0$). In this case from (5b) we obtain that $\Delta x = \dfrac{\Delta x'}{\sqrt{1-\beta^2}}$ and from (5a) that $\Delta t = \gamma\left(\dfrac{V}{c^2}\cdot \Delta x'\right) = \dfrac{V}{c^2}\cdot \Delta x$. Therefore the events aren't simultaneous anymore when seen by the observers in the stationary frame. Again we notice that the deviation from simultaneity $\Delta t$ is proportional to the apparent distance between events $\Delta x$ and we have further $\dfrac{\Delta x}{\Delta t} = \dfrac{c^2}{V} = V_f$. Obviously $V \cdot V_f = c^2$ and by analogy to the group and phase velocities of energy waves propagating through matter or to de Brogile waves[3] associated with moving particles[4,5] we interpret $V_f$ as being the phase velocity of the moving frame. A high $V_f$ means that a very small inter-frame delay or simultaneity alteration occurs due to relative movement of frames. It is also important to notice here that $V_f$ is also the phase velocity of the de Broglie waves associated by the observers in stationary frame to objects commoving with the moving frame thus moving with group velocity $V$ relative to these observers.

### 5. The Lorentz transformations and relativity of simultaneity

In the paragraph above we found that the simultaneity error depends on the distance between events and the new defined phase velocity of the moving frame $V_f = \dfrac{c^2}{V}$. Therefore the time profile is different in the two frames.

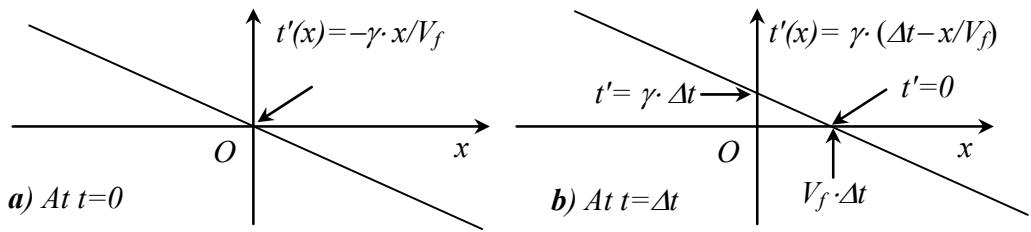

**Figure 2**. *The time profile in frame K' as occurring to observers in frame K*

For observers in one frame it occurs that the time profile in the other frame flows like a wave with finite but supra-luminal velocity $V_f$ thus causing simultaneity errors. Figure 2a above depicts the apparent time profile for the corresponding observers as predicted by the Lorentz equation (1a) just at the origin of time $t = 0$ in the stationary frame. We notice that only the corresponding observers at the origin $O,O'$ ($x = x' = 0$) read the same time. All other corresponding observers in frame K' read a different time. Figure 2b depicts this situation after a time interval $\Delta t$. The observer in K' reading $t' = 0$ has moved with velocity $V_f$ in respect to frame K. Actually the whole time profile of frame K' moves with this velocity. Therefore $V_f$ occurs as a phase velocity for the time profile (time wave) to the observers in frame K.

Because the time as perceived by common sense is not relativistic invariant we need another reference to relate the events when perceived by observers moving at different speeds.

### 6. The Lorentz transformations and the de Broglie waves

In 1923 Louis de Broglie first hypothesized that any moving particle behaves in certain circumstances like a wave[3]. Later on this was also proved experimentally. The frequency of the de Broglie wave is given by the total energy of the particle including its rest energy as predicted by Einstein's equation. By analogy to the definition for the relativistic photon energy ($E = h \cdot \omega / 2\pi$) we have for a particle moving with speed $v$:

$$\omega = 2\pi \frac{E}{h} = \frac{m_0 c^2}{\hbar \sqrt{1 - \frac{v^2}{c^2}}} \quad \text{or simplified} \quad \omega = \gamma \frac{m_0 c^2}{\hbar} = \gamma \omega_0 \tag{10}$$

The phase velocity of this wave $v_f$ depends on the particle impulse. By definition the wave number is $k = \omega / v_f$ and using the equations for the relativistic impulse:

$$p = \frac{m_0 v}{\sqrt{1 - \frac{v^2}{c^2}}} \quad \text{and relativistic wave number:} \quad p = \hbar k \tag{11}$$

and the result in (10) we get:

$$v_f = \frac{k}{\omega} = \frac{c^2}{v} \tag{12}$$

Therefore de Broglie associates with any moving particle a relativistic invariant combination $v \cdot v_f = c^2$ as the product between the particle velocity and the phase velocity of its de Broglie wave. It is interesting to note that the wave associated with the moving particle behaves like a wave propagating through a medium with normal dispersion $v < c < v_f$. The wave equation may be written as:

$$A(x,t) = A_{max}(x,t) \cdot \cos\omega\left(t - \frac{x}{v_f}\right) \tag{13}$$

In this context the wave amplitude has a special significance only in relation to quantum mechanics theory and it gives the probability to localise the particle or its energy at certain time $t$ and at certain point $x$ in space. Therefore by this approach the particle speed $v$ represents the group velocity of the energy transported by its de Broglie associated wave.

We extend the de Broglie hypothesis also to stationary particles. In this case the frequency of the associated stationary wave is given by the particle rest energy as $\omega_0 = \frac{m_0 c^2}{\hbar}$ and its phase velocity given by (12) becomes infinite $v_f \rightarrow \infty$. The group velocity of this wave is zero $v=0$ as the particle or its associated wave doesn't transport any energy, thus the wave equation becomes:

$$A(x',t') = A_{max}(x') \cdot \cos\omega_0 t' \tag{14}$$

Now consider the same particle as seen from two inertial reference frames:
1. a first stationary frame K in respect thereof our particle moves with constant speed $v$
2. a second frame K' commoving with this particle.

Observers in the stationary frame associate with the moving particle a de Broglie wave using equation (13) whereas the observers in the frame commoving with the particle associate with the same particle a stationary wave given by the equation (14). Imposing the condition that the phase of these waves is the same for all corresponding observers we get the equation:

$$\omega_0 t' = \omega\left(t - \frac{x}{v_f}\right) \tag{15}$$

From (10) we have $\omega = \gamma \omega_0$ and because $v_f = c^2/v$ the result from (15) reduces to the well-known Lorentz transformation of time $t' = \gamma\left(t - \frac{x}{v_f}\right)$. We conclude that the phase of the de Broglie wave is a relativistic invariant. We may possibly associate a wave to any particle that encloses or transports energy at sub-luminal speeds. Thus we could interpret the space-time universe and its matter bounded energy as being a superposition of de Broglie waves such that the phase of each wave is invariant for all corresponding observers in any frame.

## 7. The Lorentz transformations and the light waves

Albert Einstein first revealed in 1905 the particle behaviour of light in respect to the photo-electric phenomenon. Photons are energy particles transporting energy in a non-dissipative medium at group velocity $v = c$ and phase velocity $v_f = c$. Therefore consistently we also have $v \cdot v_f = c^2$. For the sake of generalisation we admit that the light seen as an electromagnetic wave represents the de Broglie wave associated with the photon seen as a particle transporting energy. In a non-dissipative medium the wave energy density is constant, which means that the wave amplitude is constant and by the quantum mechanics approach it means that the probability to find the wave energy / the photons is the same at any point within the wave space. We are now mainly concerned with the phase of this electromagnetic wave. We start with the Lorentz equations in form (1) and we divide (1b) by $c$ and subtract the result term by term from (1a) to obtain:

$$t' - \frac{x'}{c'} = \sqrt{\frac{1+\beta}{1-\beta}} \left( t - \frac{x}{c} \right) \qquad (16)$$

By a similar procedure we obtain that:

$$t' + \frac{x'}{c'} = \sqrt{\frac{1-\beta}{1+\beta}} \left( t + \frac{x}{c} \right) \qquad (17)$$

The wave equations for an electromagnetic wave propagating in the positive direction of OX axis as seen from the two reference inertial frames are:

$$E(x,t) = E_{max} \sin \omega \left( t - \frac{x}{c} \right) \qquad B(x,t) = B_{max} \sin \omega \left( t - \frac{x}{c} \right) \qquad (18)$$

$$E'(x',t') = E'_{max} \sin \omega' \left( t' - \frac{x'}{c} \right) \qquad B'(x',t') = B'_{max} \sin \omega' \left( t' - \frac{x'}{c} \right)$$

Imposing the condition that the wave phase is the same for all corresponding observers at any time and at any point in space we get:

$$\omega' \left( t' - \frac{x'}{c} \right) = \omega \left( t - \frac{x}{c} \right) \qquad (19)$$

In this relation we replace the result (16) that we found above and we obtain finally that $\omega' = \omega \sqrt{\frac{1-\beta}{1+\beta}}$, which is the known and well-founded relativistic Doppler effect. We obtain the same result for the waves propagating in the negative direction of the OX axes by using the intermediate result in (17). Therefore we conclude that beside the speed of light, the phase of the electromagnetic wave is also a relativistic invariant being the same for all corresponding observers in any frame.

### 8. More about the wave-particle duality

Physical entities without rest mass as the photons and the neutrinos propagate in vacuum transporting energy at light speed. We conclude that only physical entities without rest mass $m_0=0$ would transport a finite amount of energy at light speed. Otherwise with $v \rightarrow c$ the famous relation $E = \dfrac{m_0 c^2}{\sqrt{1-\dfrac{v^2}{c^2}}}$ predicts an infinite amount of energy. Particles with non-zero rest mass transport energy at sub-luminal speed $v<c$, whereas the phase velocity of the de Broglie wave associated with the moving particle has supra-luminal values $v_f>c$ such that their product remains constant and invariable $v \cdot v_f = c^2$.[4,5] Now, consider the electromagnetic wave propagating in fundamental mode $TE_{10}$ through a hollow microwave conductor also known as a waveguide. In this case the wave energy propagates through the waveguide with sub-luminal group velocity $u$:

$$v = c\sqrt{1-\left(\dfrac{\omega_0}{\omega}\right)^2} < c \qquad (20)$$

and with supra-luminal phase velocity $w$:

$$v_f = \dfrac{c}{\sqrt{1-\left(\dfrac{\omega_0}{\omega}\right)^2}} > c$$

(21)

where $\omega_0$ is the "cutoff frequency" of the waveguide. We still have $v \cdot v_f = c^2$ but in this case the values of group and the phase velocities are speaking rather for the particle-like and less for the wave-like energy propagation. Therefore we conclude that the electromagnetic wave propagating through the waveguide (matter bounded energy) behaves mainly like a particle with rest mass. We can even calculate the rest mass of the particle associated with the matter bounded photon. With $k = \omega/v_f$ we get:

$$m_0 = \dfrac{\hbar \omega_0}{c^2}. \qquad (22)$$

Surprisingly it depends only on waveguide parameters and doesn't depend on wave frequency.

Furthermore, the same rationale also holds for the light propagating through a transparent medium with normal dispersion, thus having the refractive index $n > 1$. In this case the group velocity decreases to sub-luminal values and the light propagation likely approaches the particle aspect. As above we calculate the rest mass of the particle equivalent to the matter bounded light photon with:

$$m_0 = \frac{hf\sqrt{1-\frac{1}{n^2}}}{c^2}. \qquad (23)$$

We speculate that the particle aspect of light propagation dominates when the light propagates through or interacts with the matter.

### 9. Conclusions

Based on Lorentz transformations we have learned that whilst the time and space in the special relativity ceases to be absolute and invariant physical entities, the phase of the de Broglie waves associated with moving matter or energy may replace our usual representation of time or space as the absolute reference for all observers.

For the quantum mechanics theory the amplitude of the de Broglie wave seems to be of main interest by its association with the probability to find the particle and energy at some time and place in universe. At contrary for the special relativity theory the phase of this wave is of main interest as it connects the space and time coordinates of different observers together.